# Giant anomalous Hall effect in ultrathin Si/Fe bilayers

S. S. Das and M. Senthil Kumar[*]

*Department of Physics, Indian Institute of Technology Bombay, Mumbai 400076, India*

**Highlights**

- The AHE studies on the ultrathin Si/Fe bilayers shows significant enhancement of about 60 times in the saturation anomalous Hall resistivity ($\rho_{hs}^A$) and about 265 times in the anomalous Hall coefficient ($R_s$).

- The largest value of $\rho_{hs}^A = 2.6 \times 10^{-7}$ Ω m observed is about three orders of magnitude larger than that reported for pure Fe.

- The largest value of anomalous Hall sensitivity S = 433 Ω/T observed for $t_{Fe}$ = 10 Å surpasses that of the semiconducting Hall sensors.

- The scaling law between the AHE and longitudinal electrical resistivity ($\rho$) suggests side jump as the dominant mechanism of AHE.

**Abstract**

Anomalous Hall effect studies on ultrathin Si(50Å)/Fe($t_{Fe}$) bilayers were performed at 300 K. Giant enhancements of about 60 times in saturation anomalous Hall resistivity and 265 times in anomalous Hall coefficient ($R_s$) were observed upon decreasing the Fe layer thickness $t_{Fe}$ from 200 to 10 Å. The $R_s$ observed for $t_{Fe}$ = 10 Å is about three orders of magnitude larger than that of bulk Fe. Scaling law between $R_s$ and longitudinal electrical resistivity ($\rho$) suggests that the side jump is the dominant mechanism of the anomalous Hall effect. The observed largest Hall sensitivity of 433 Ω/T surpasses that of the semiconducting GaAs and InAs Hall sensors already reported.

[*]Corresponding author (email: senthil@iitb.ac.in)

# 1. Introduction

Currently, there is a growing research interest on ferromagnetic heterostructures exhibiting interesting spin-dependent transport properties such as anisotropic magnetoresistance, giant magnetoresistance, and tunneling magnetoresistance. Besides these properties, there is another spin-dependent phenomenon called the anomalous Hall effect (AHE) or the extraordinary Hall effect (EHE) [1]. The AHE which is observed in magnetic materials now extends its application possibilities in the magnetic characterization of nanostructures, thin films, recording layer of double-layered perpendicular magnetic recording media, etc. [2, 3]. Recently, Lu *et al.* have reported the highest anomalous Hall sensitivity of 12000 $\Omega$/T in $SiO_2/FePt/SiO_2$ sandwich structure film. This is about an order of magnitude larger than the best semiconductor sensitivity reported so far [4]. High sensitivity, weak temperature dependence, linear field response, hysteresis-free behavior with the field, and simple fabrication process of the AHE based devices pioneer new possibilities in the field of magnetic sensors.

Despite a lot of research carried out on the AHE since its discovery [5], the effect still remains less understood and hence demands more detailed study. Furthermore, the smaller Hall coefficient and Hall sensitivity of bulk magnetic materials hinder their application possibilities. Many efforts have been made to achieve the enhancement in the AHE and accordingly the effect has been studied in various layered structures such as Fe/Cu, Co/Pd, Fe/Cr, Fe/Ge, Co/Cu, etc. [6-11]. However, not much studies have been carried out on the Si/Fe system. In our previous papers on the Si/Fe multilayers [12, 13], we have reported the important role of the Si-Fe interfaces in enhancing the AHE. As a continuation of our research activities on the Si/Fe system, in this letter, we report the study extended by us on the ultrathin Si/Fe bilayers.

## 2. Experimental Details

Si(50 Å)/Fe($t_{Fe}$) bilayers were deposited onto glass and silicon substrates simultaneously by dc magnetron sputtering at an Ar pressure of $4 \times 10^{-3}$ mbar. The base vacuum achieved prior to the deposition was $2 \times 10^{-6}$ mbar. A power of 40 W was used for sputtering of both the Fe and Si targets. The nominal thickness of the Fe layer ($t_{Fe}$) was varied from 10 to 200 Å whereas that of the Si layer was kept fixed at 50 Å which corresponds to the maximum AHE in the Si/Fe multilayers as reported by us [12]. More details of the deposition conditions were reported by us elsewhere [12]. Hall effect measurements were performed at 300 K by circular four-probe method with the magnetic field varying up to 27 kOe.

## 3. Results and discussion

For magnetic thin films, the empirical expression for Hall resistivity ($\rho_h$) can be written as [9, 14-16],

$$\rho_h = R_h \cdot t = R_0 H + R_s 4\pi M \tag{1}$$

where $R_h$, $t$, $R_0$, $R_s$, and $M$ are anomalous Hall resistance, thickness of the bilayer, ordinary Hall coefficient, anomalous Hall coefficient, and magnetization, respectively. In the RHS of eq. (1), the first term corresponds to the ordinary Hall effect (OHE) arising because of the Lorentz force on the charge carriers and the second term represents the AHE resulting from the break of the right-left symmetry during spin-orbit scattering of electrons. The $R_s$, saturation anomalous Hall resistivity $\rho_{hs}^A$ and perpendicular saturation field $H_s$ can be calculated from the Hall effect data by following the procedures described in Refs. 10, 14 and 17. The $\rho_{hs}^A$ can be obtained by extrapolating the $\rho_h$ versus H data to H = 0. The $H_s$ was taken as the field at which $\rho_h$ begins to saturate. Since both the AHE and the longitudinal electrical resistivity $\rho$ have their origin from a scattering event, the $R_s$ in eq. (1) and $\rho$ follow the scaling law, [6, 11, 17, 18]

$$R_s = a\rho + b\rho^2, \tag{2}$$

where the first term represents the skew scattering and the second term corresponds to the side jump contribution. The values of the coefficients $a$ and $b$ provide information about the dominant mechanism of the AHE. Besides the skew scattering and side jump mechanisms which have an extrinsic origin, there is another intrinsic contribution to AHE due to the Berry phase or Berry curvature [15, 16]. However, in the case of our heterogeneous Si/Fe bilayers, the intrinsic contribution to the AHE is negligible due to the dominant role of interface roughness/interdiffusion present at the interfaces that results in large magnetic disorder.

Fig. 1a shows the $\rho_h$ versus $H$ plot of the Si(50Å)/Fe($t_{Fe}$) bilayer samples at 300 K. In the low field regime, this graph shows linear behaviour with negligible hysteresis. This suggests that the easy direction of magnetization lies in the plane of the film. This in-plane magnetic anisotropy has also been confirmed from the in-plane and perpendicular magnetization loops (not shown) of the samples. The positive $R_0$ obtained from the slope of the high field regime Hall data confirms the hole-type conduction in the samples. From the analysis of the AHE data of all the samples, the $\rho_{hs}^A$ and $R_s$ have been obtained and they were normalized with respect to that for $t_{Fe}$ = 200 Å. These normalized data are plotted as a function of $t_{Fe}$ in Fig. 1b. Giant enhancements of about 60 times in $\rho_{hs}^A$ and about 265 times in $R_s$ have been obtained upon decreasing the $t_{Fe}$ from 200 to 10 Å. This enhancement in both $\rho_{hs}^{A}$ and $R_s$ is very large when compared with Fe and Ni single films. An increase of about 3-4 times in $\rho_{hs}^{A}$ and $R_s$ has been reported for the Fe single-layer films upon decreasing $t_{Fe}$ from 1700 to 80 Å [19]. Similarly, an increase of about three times in $\rho_{hs}^{A}$ and about four times in $R_s$ has been reported in the case of Ni single film upon decreasing the film thickness from 200 to 20 Å [20]. A change of about two times in the AHE was reported in the ultrasensitive $SiO_2$/Fe-Pt/$SiO_2$ sandwich structures within the thickness range of 14-200 Å [4]. The largest value of $\rho_{hs}^{A}$ = 2.6 × $10^{-7}$ Ω m observed for $t_{Fe}$ = 10 Å in our Si/Fe bilayers is about three orders of magnitude larger than $\rho_{hs}^{A}$ = $10^{-10}$ Ω m reported for pure Fe [21] of thickness 0.5 μm. Similarly, $R_s$ = 5.1 × $10^{-7}$ Ω m/T for $t_{Fe}$ = 10 Å is about three orders of magnitude larger than (2.8-7.2) × $10^{-10}$ Ω m/T reported for the bulk Fe [9, 15]. This large enhancement in $\rho_{hs}^{A}$ and $R_s$ of our samples may be due to an increase of surface and interface scattering in the ultrathin layers. The Si/Fe interface plays a major role in the enhancement of the AHE by increasing the interface scattering and simultaneously reducing the short circuit and shunting effects, which are essentially observed in multilayers with a conducting spacer [6]. In our earlier paper, we have investigated the AHE by varying the number of bilayers (N) of $[Si/Fe]_N$ multilayers [13]. In this case, the $\rho_{hs}^{A}$ increases by three times from 9.0 × $10^{-8}$ Ω m to 2.6 × $10^{-7}$ Ω m upon decreasing N from 20 to 1. Such a large AHE of the bilayers as compared to that of the multilayers is due to the large magnetic disorder at the interfaces for N = 1. Thus, the scattering of the Hall carriers due to the surface/interface disorders is more prominent in bilayers as compared to that of the multilayers.

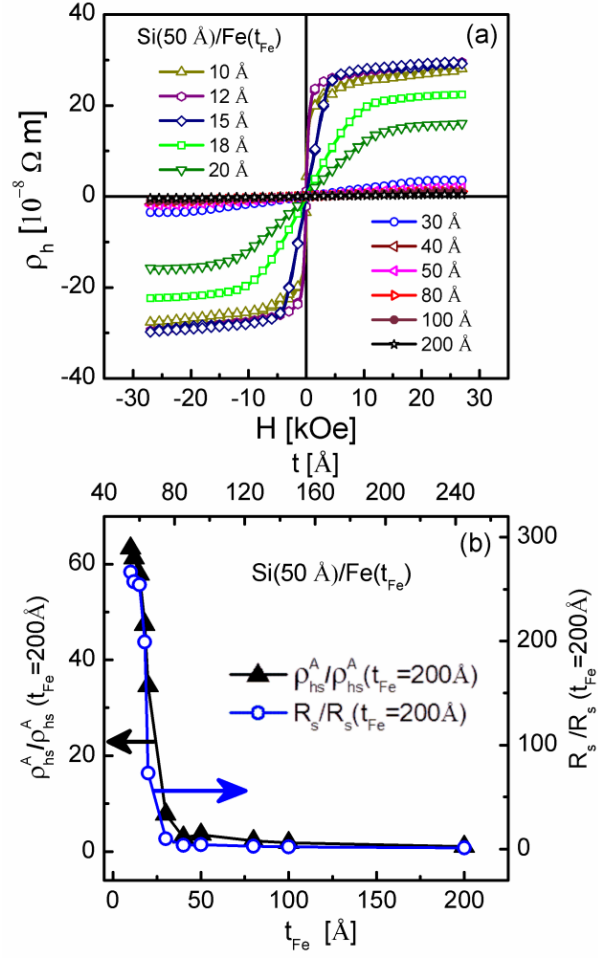

Fig. 1 (a) The $\rho_h$ versus H graph for the Si(50Å)/Fe($t_{Fe}$) bilayers at 300 K. The data points other than $t_{Fe}$ = 18 and 20 Å lie very close to each other. (b) The $t_{Fe}$ dependence of $\rho_{hs}^A$ and $R_s$ normalized with respect to that for $t_{Fe}$=200 Å. Total thickness $t$ of the bilayers is plotted in the top x-axis for convenience.

To understand the mechanism of the AHE in the Si/Fe bilayers, we have verified the scaling law between $R_s$ and $\rho$ using eq. (2). The values of the coefficients $a$ and $b$ can be obtained from the linear fit of the graph between $R_s/\rho$ versus $\rho$ as shown in Fig. 2. The coefficients $a$ and $b$ obtained from the above fitting are $(3.27 \pm 0.60) \times 10^{-3}$ T$^{-1}$ and $(6.54 \pm 0.12) \times 10^3$ Ω$^{-1}$ m$^{-1}$ T$^{-1}$, respectively. As $b$ is about six orders of magnitude larger than that of $a$ it is clear that the side jump mechanism is the dominant mechanism in the Si/Fe bilayers. Similar studies in other systems such as Co/Pd multilayers [11], MnBi films [22], etc. have also been reported and the larger magnitude of $b$ over $a$ is also seen in these cases. The $\rho$ of our bilayer samples measured at $H = 0$ shows an increase (~25 times) from 3.5 × 10⁻⁷ Ω m to 8.7 × 10⁻⁶ Ω m when $t_{Fe}$ decreases from 200 to 10 Å. This large increase of the resistivity can be understood as the effect of the increase in the discontinuity of the Fe layer with the decrease of $t_{Fe}$ as observed from the HRTEM data (not shown) of the samples. Further, XRD data of the samples (not shown) shows a significant decrease in grainsize from about 105 Å to 13 Å upon decreasing $t_{Fe}$ from 200 Å to 10 Å resulting in the increase of surface-to-volume ratio and hence the increased surface/interface scattering.

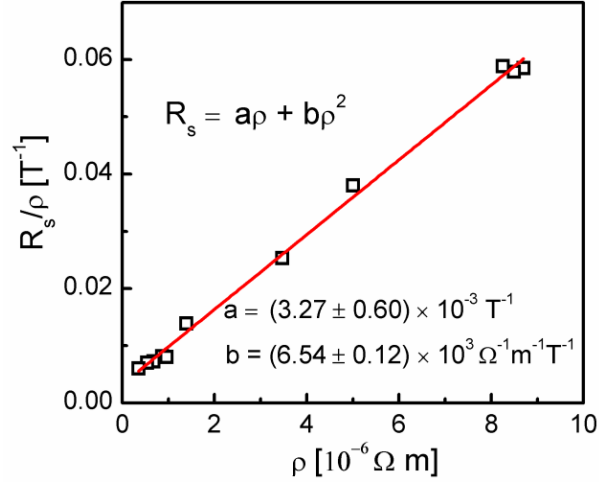

Fig. 2 The $R_s/\rho$ versus $\rho$ plot to verify the scaling law between $R_s$ and $\rho$ for the Si/Fe bilayers. Here $\rho$ was measured at $H = 0$. The open squares are the data points and the solid line is the fit using eq. (2).

The Hall sensitivity ($S$) of the bilayers is plotted as a function of $t_{Fe}$ in Fig. 3. An increase of $S$ from 0.08 to 433 Ω/T has been observed upon decreasing $t_{Fe}$ from 200 to 10 Å. This large enhancement in $S$ is due to the increase of $R_s$ as shown in Fig. 1b and it is also due to the decrease of the perpendicular saturation field $H_s$. The $t_{Fe}$ dependence of $H_s$ which is also depicted in Fig. 3 shows the decrease of $H_s$ from 22.6 to 0.62 kOe when $t_{Fe}$ decreases from 200 to 10 Å. Similar decrease of $H_s$ in the ultrathin films has also been reported by some authors [1, 2]. **For** comparison with the Si/Fe multilayers, we have also studied the effect of varying the number of bilayers (N) on the S of the [Si/Fe]$_N$ multilayers [13]. In this study, S is observed to increase from 1 to 433 Ω/T when N is decreased from 20 to 1. In the case of the bilayers, this suggests that the simultaneous decrease of $H_s$ and enhanced scattering of Hall carriers due to surface/interface disorders increases $\rho_{hs}^A$ resulting an enhanced S. The largest value of $S = 433$ Ω/T (within the field range –0.65 kOe to +0.65 kOe) obtained for $t_{Fe} = 10$ Å is very high as compared to other multilayers and metal-insulator granular systems. This is even larger than that of some semiconductors such as Si, Ge, GaAs and InAs [9, 14, 23]. A Hall sensitivity of about 190 Ω/T has been reported in the case of GaAs Hall sensor (Siemens KSY 10) [23]. Similarly, in the case of InAs Hall sensor, the sensitivity of 95 Ω/T at a temperature of 20°C has also been reported [23]. A Hall sensitivity of 82 Ω/T has also been reported in the case of FeGe amorphous composite film [14]. The largest $S$ of 12000 Ω/T has been reported in perpendicular anisotropic SiO$_2$/Fe-Pt/SiO$_2$ sandwich films [4]. However, the enhancement of $\rho_{hs}^A$ and $R_s$ observed in this system is only about 2-3 times within the thickness range 14 to 200 Å. The giant enhancement of the AHE observed in our Si/Fe bilayers clearly indicates that the sensitivity can further be enhanced if more improvement in the perpendicular anisotropy is achieved.

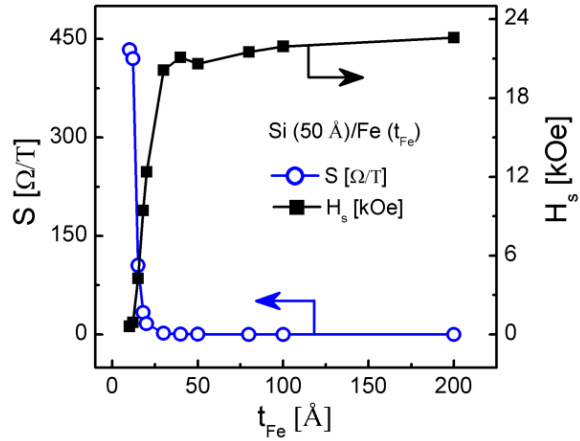

Fig. 3 The $t_{Fe}$ dependence of the Hall sensitivity ($S$) and perpendicular saturation field ($H_s$) for the Si/Fe bilayers.

## 4. Conclusions

In conclusion, we have observed giant enhancements of about 60 times in $\rho_{hs}^A$ and about 265 times in $R_s$ in the Si/Fe bilayers upon decreasing $t_{Fe}$ from 200 to 10 Å. The observed largest values of $R_s$ and $R_0$ are about three orders of magnitude larger than that of bulk Fe. The sensitivity increases by three orders of magnitude upon decreasing $t_{Fe}$ from 200 to 10 Å. The largest value of $S$ = 433 Ω/T for $t_{Fe}$ = 10 Å surpasses the OHE based semiconducting sensors. The large enhancement in the AHE of the Si/Fe bilayers envisages that the material could be a promising candidate for Hall sensors.

**References**

[1] Gerber A. J Magn Magn Mater 2007; 310:2749-51.

[2] Gerber A, Riss O. J Nanoelectron Optoelectron 2008; 3:35-43.

[3] Kumar S, Laughlin DE. IEEE Trans Magn 2005; 41:1200-08.

[4] Lu YM, Cai JW, Pan HY, Sun L. Appl Phys Lett 2012; 100:022404-4.

[5] Hall EH. Am J Math 1879; 2:287-92.

[6] Xu W, Zhang B, Wang Z, Chu S, Li W, Wu Z *et al.* Eur Phys J B 2008; 65:233-37.

[7] Shaya O, Karpovski M, Gerber A. J Appl Phys 2007; 102:043910-5.

[8] Song SN, Sellers C, Ketterson JB. Appl Phys Lett 1991; 59:479-81.

[9] Liu YW, Mi WB, Jiang EY, Bai HL. J Appl Phys 2007; 102:063712-7.

[10] Tsui F, Chen B, Barlett D, Clarke R, Uher C. Phys Rev Lett 1994; 72:740-43.

[11] Guo ZB, Mi WB, Aboljadayel RO, Zhang B, Zhang Q, Barba PG *et al.* Phys Rev B 2012; 86:104433.

[12] Das SS, Senthil Kumar M. J Phys D: Appl Phys 2013; 46:375003.

[13] Das SS, Senthil Kumar M. IEEE Trans Magn 2014; 50:2005604.

[14] Liu H, Zheng RK, Zhang XX. J Appl Phys 2005; 98:086105-3.

[15] O'Handley RC. Modern Magnetic Materials Principles and Applications New York: Wiley; 1999.

[16] Hurd CM. The Hall Effect in Metals and Alloys. New York: Plenum Press; 1972.

[17] Smit J. Physica 1955; 21:877-87.

[18] Berger L. Phys Rev B 1970; 2:4559-66.

[19] Galepov PS. Soviet Physics Journal 1969; 12:133-35.

[20] Volkov V, Levashov V, Matveev V, Matveeva L, Khodos I, Kasumov Y. Thin Solid Films 2011; 519:4329-33.

[21] Zhao B, Yan X. J Appl Phys 1997; 81:4290-92.

[22] Kharel P, Sellmyer DJ. J Phys: Condens Matter 2011; 23:426001.

[23] Heremans J. J Phys D: Appl Phys 1993; 26:1149.

**Figure Captions:**

Fig. 1 (a) The $\rho_h$ versus H graph for the Si(50Å)/Fe($t_{Fe}$) bilayers at 300 K. The data points other than $t_{Fe}$ = 18 and 20 Å lie very close to each other. (b) The $t_{Fe}$ dependence of $\rho_{hs}^A$ and $R_s$ normalized with respect to that for $t_{Fe}$=200 Å. Total thickness $t$ of the bilayers is plotted in the top $x$-axis for convenience.

Fig. 2 The $R_s/\rho$ versus $\rho$ plot to verify the scaling law between $R_s$ and $\rho$ for the Si/Fe bilayers. Here $\rho$ was measured at $H$ = 0. The open squares are the data points and the solid line is the fit using eq. (2).

Fig. 3 The $t_{Fe}$ dependence of the Hall sensitivity ($S$) and perpendicular saturation field ($H_s$) for the Si/Fe bilayers.